# Silicon clathrates for photovoltaics predicted by a two-step crystal structure search

Juefei Wu, Hao Gao, Kang Xia, Dingyu Xing, Jian Sun*

National Laboratory of Solid State Microstructures, School of Physics and Collaborative Innovation Center of Advanced Microstructures, Nanjing University, Nanjing 210093, China.

**Abstract**

Silicon in a cubic diamond structure currently plays a significant role in the photovoltaic industry. However, the intrinsic band structures of crystalline silicon restrict its sunlight conversion efficiency. Recently, a clathrate-like Si-24 has been successfully synthesized, which has a quasi-direct bandgap and sheds light on silicon-based photovoltaics. Here, we proposed a two-step crystal structure search method based on first-principles calculations and explored silicon clathrate structures extensively. First, the guest-host compounds were searched at high pressure, and then, the porous guest-free silicon clathrates were obtained by removing the guest atoms. Using potassium as the guest atom, we identified four metastable silicon clathrate structures, and some of them have bandgaps close to the optimal range of the Shockley-Queisser limit and have a better absorption rate than the cubic diamond silicon. These silicon clathrates may have promising value in photovoltaic applications.

Corresponding author: J.S. (jiansun@nju.edu.cn)

Among various energy sources capable of producing clean power, solar energy plays a unique role in sustainable energy production[1]. Currently, the most widely used photovoltaic devices are made of cubic diamond silicon (D-Si). However, the intrinsic properties of D-Si, e.g., the indirect bandgap (1.1 eV) and large direct optical bandgap (3.4 eV), limit its absorption rate of sunlight. Silicon based semiconductors have their own advantages, including low cost, availability to current integrated circuit technology, non-toxicity, and environmental friendliness. Compared with silicon, compounds such as gallium arsenide (GaAs), indium phosphide (InP) and gallium nitride (GaN) are more expensive and complicated. In addition, researchers have developed several techniques to improve the sunlight emission (absorption) efficiency for silicon[2,3], such as plasmonics[4] and nanostructures[5]. Therefore, silicon based materials still receive tremendous attention.

Silicon exhibits abundant phase transitions under high pressure. For instance, D-Si transforms into a $\beta$-Sn structure at around 12GPa[6], then a primitive hexagonal structure at 42GPa[7], and a face centered cubic at 78GPa[8]. Accompanied by these phase transitions, the electronic properties of silicon alter from indirect-gap semiconducting to metallic or even direct-gap semiconducting[9]. Considering the abundant phase transitions and various electronic properties, it should be possible to find new silicon structures with properties suitable for photovoltaic applications. Apart from the bulk phase transitions mentioned above, group IV elements can also form open frameworks resembling clathrate hydrates[10]. Among various structures, clathrates with the formula $M_8X_{46}$ (clathrate-1) and $M_{24}X_{136}$ (clathrate-II), particularly those based on Si and Ge, have drawn much attention. Some compounds exhibit superconductivity[11], while others possess excellent thermoelectric properties[12]. In addition, the guest-free semiconducting clathrates, such as $Si_{46}$ and $Si_{136}$, have direct band gaps[13], suggesting their great potential for future photovoltaic applications.

To explore silicon structures fit for solar absorptions, several crystal structures have been proposed using first principles calculations[14-21]. Some of these predicted metastable silicon structures have direct or quasi-direct band gaps within the Shockley-Queisser limit[22,23] and belong to the clathrate family. Meanwhile, experimentalists have produced a guest-free porous silicon clathrate-like structure by evaporating the sodium guest atoms from a $Na_4Si_{24}$ precursor synthesized under high pressure and at high temperature[24]. The resultant pure silicon allotrope is an orthorhombic structure (space group *Cmcm*, no. 63) and named as Si-24. It has a quasi-direct bandgap that overlaps well with the sunlight spectrum. Inspired by this, we propose to synthesize silicon clathrates with different tunnel size by using different alkali metal based on a two-step method. First, guest atoms with different sizes are added and stable guest-host compounds are searched under pressure; second, the guest atoms are removed and possible porous candidates are picked out. As for the guest atoms, potassium is widely used in the clathrate compound synthesis. Besides, potassium also belongs to the IA group, which may have similar properties as sodium in the precursors but with a different atomic size. Thus, we use potassium atoms as the first example to search possible potassium silicon compounds.

In this letter, the silicon clathrate structures were produced by applying random search algorithm[25,26] together with first-principle calculations. During the structure searches, compositions between potassium and silicon are 1:4, 1:5, 1:6 and 1:8. We chose these proportions based on the following reasons: first, silicon clathrate compounds are supposed to be silicon rich; second, these compositions are around the proportion of the precursor ($Na_4Si_{24}$) for Si-24 and other classical silicon clathrate compounds[12]. The pressure for structure search calculations is 20 GPa. Although the pressure is higher than that to synthesize the Si-24 precursor, it can enhance the possibility to overcome barriers and find stable structures during simulations. *Ab initio* calculations for energetics and structure relaxations were performed using projector augmented wave (PAW) formalism as implemented in the VASP[27,28] package, together with the Perdew-Burke-Erzernhof (PBE) generalized gradient approximation (GGA) exchange-correlation functionals[29]. We set the energy cutoff for the plane wave basis to be 390 eV and 370 eV for structures with and without guest atoms, respectively. All forces are converged to be better than 0.003 eV/Å. The Brillouin zone was sampled by Monkhorst-Pack meshes with a k-spacing of 0.025 /Å in order to provide sufficient accuracy during enthalpy and phonon calculations. We verified the dynamic stability of the structures by calculating the phonon spectra using the direct supercell approach as implemented in the PHONOPY[30]. $2 \times 2 \times 2$ Supercells were used for phonon calculations. Since calculations with the conventional PBE functionals systematically underestimate bandgaps, we studied the electronic and optical properties using the modified Becke-Johnson exchange potential (mBJ)[31] + PBE-correlation implanted in WIEN2K[32], which produces bandgaps with an accuracy similar to very expensive GW calculations.

We calculated the convex hull to check the formation enthalpy of every compound structure. For every composition, we discarded the compound structures with energies 150 meV/atom higher than the convex hull curve. Subsequently, we remove the potassium atoms from the compounds, and the bandgaps of the pure silicon structures were calculated at 0 GPa. If the total energy of the structure is very low, or the structure has direct/quasi-direct bandgap and the gap lies within 0.5-2.0 eV, the silicon structure will be picked out and rechecked with more accurate criteria. After this process, we picked out four best structures from thousands of candidates. These structures are displayed in Fig. 1 (a), and their detailed lattice information is given in the supplementary material. The formation enthalpy ($H_f$) of the compounds under high pressure is defined as

$$H_f = E(K_m Si_n) - mE(K) - nE(Si) \quad (1)$$

As shown in Table 1, $H_f$ of all the four compounds is negative, suggesting that these compounds are energetically stable and have good chance to be synthesized.

The enthalpies of the proposed silicon clathrate structures are compared with those of the diamond silicon, clathrate-I (Si-46), the synthesized Si-24 and previously predicted $Si_{20}$-T[15]. As depicted in the Fig. 1 (b), diamond silicon remains to be the most stable one at the pressure range of 0-10 GPa. Silicon structures proposed in this work are more energetically favorable than the $Si_{20}$-T structure at zero pressure, suggesting that the predicted silicon structures may be metastable. The large energy drop of some curves for the pure silicon structure after 10GPa is due to the large distortion of the porous structures. But all these porous silicon structures can keep their configuration when the pressure is under 10 GPa. Among the proposed silicon structures, energies of *Cmmm*-24 and *Cmcm*-24 are comparable to that of Si-24 at zero pressure. In particular, the energy difference between *Cmmm*-24 and Si-24 is less than 1 meV, which is almost within the error of DFT; while *Cmcm*-24 is about 0.025 eV/atom above Si-24. Although the difference between *Cmmm*-24 and Si-24 becomes larger with the increasing in pressure and Si-24 is more favorable, *Cmmm*-24 is still comparable to Clathrate-I structure (Si-46) and becomes better than Si-46 when the pressure is above 5GPa. Therefore, according to our calculations discussed above, the silicon structures we predicted in this work are possible to be synthesized under proper conditions.

Usually, the polygon units in the silicon structures and their distortions have direct relations to the strain energies. According to the calculations by Karttunen *et al.*[33], five- and six- membered rings are more favorable, while four-membered rings are much more strained. After removing the guest potassium atoms, our predicted silicon structures contain tunnel-like voids consisting of six-, eight- or ten-membered rings. The *Cmmm*-16 structure contains a four-membered ring, a distorted six-membered ring, and eight-membered ring from different observing directions, which appears analogous to bct-carbon[34] and tI16-Si[16]. In comparison, the other three structures have layers or frameworks constructed with edge shared five-membered rings, similar to that in the M-carbon[35,36]. Silicon atoms in these four silicon structures are mostly $sp^3$-like 4-coordinated except some atoms in the *C2/m*-16 structure. Combining our results with some other reported clathrate structures, such as the recently synthesized Si-24, it suggests that the edge shared five-membered unit plays an important role in the silicon clathrate structures. This is in agreement with the regularity that the five-membered ring is less strained compared with other configurations, especially the four-membered ring[33]. Under zero pressure, the sequence of the predicted silicon structures starting from the most favored is *Cmmm*-24, *Cmcm*-24, *C2/m*-16 and *Cmmm*-16. As depicted in Fig. 1 (a), *Cmcm*-24 has a high ratio of the energetically favorable five-membered rings, while *Cmmm*-24 is composed of five- and six-membered rings. Therefore, *Cmmm*-24 and *Cmcm*-24 have a lower enthalpy than *C2/m*-16 and *Cmmm*-16 under ambient pressure. Although structure *Cmmm*-24 has the component of the four-membered ring, the distorted six-membered ring makes it relatively favorable than others.

Then, we examined the dynamical stability of the potassium silicon compounds and their corresponding guest-free clathrate structures following the synthetization process of Si-24, namely, to synthesis the precursor at first, then recover the precursor to the ambient condition, and finally vaporize the guest atoms out from the compounds. The phonon spectra of all the proposed silicon structures at ambient pressure are displayed in Fig 2, proving that all of them are dynamically stable. According to our calculations, the *Cmmm*-24 and *Cmmm*-16 compounds are also stable at high pressures; the results are shown in the Supplementary Material (Fig. S1). In addition, after quenching to the ambient pressure, the four proposed structures with guest atoms retain their configuration, and phonon spectra of the compounds (in Supplementary Fig.S2) have no imaginary frequency at ambient pressure.

Since Si-24[24] and other porous silicon structures[14-21] have potential applications in photovoltaics, it should be interesting to look into the electronic and optical properties of these newly predicted silicon clathrate structures. The bandgap is one of the most crucial parameters to determine the conversion efficiency of photovoltaic materials. According to the Shockley-Queisser limit, materials with a gap around 1.3 eV are superior to reach high conversion efficiency[22,23]. Because calculations with conventional PBE functionals usually underestimate the bulk gaps, we applied the mBJ potentials to calculate the bulk gaps. The precision of mBJ for calculating the bandgap is comparable with GW method and hybrid functional calculations (such as HSE06). But GW and HSE06 calculations are much more expensive as cells of our predicted silicon structures in this work are rather big. The calculated bandgaps of the four predicted silicon clathrates are listed in Table II and compared with that of diamond silicon. In particular, the calculated bandgap of diamond silicon is in good agreement with the experiments, showing the reliability of our calculation method. Fig. 3 demonstrates the electronic band structure of the proposed silicon structures. Except the *C2m*-16 structure, the other three silicon structures are indirect bandgap semiconductors. The bandgaps of the three structures are within the optimal range of the Shockley-Queisser limit. Among the proposed silicon structures, the direct bandgap of *Cmmm*-24 is around 1.4 eV, which is the closest to the 1.3 eV limit. The VBM at the Y point and CBM at G make *Cmcm*-24 the indirect bandgap material. Direct bandgaps of *Cmcm*-24 at G and Y are close to each other, which are 1.02 eV and 1.1 eV, respectively. Therefore, the conversion efficiency of this material could be similar to that of direct gap materials. Meanwhile, some methods such as strain might be used to tune its band dispersion to achieve an even better performance. Another interesting phenomenon is that the *C2/m*-16 seems to be a semimetal and has a tiny gap at the Z point, around 35 meV. There are very large dispersions in the band structures of *C2/m*-16, indicating carriers in this structure should have pretty large mobility.

Followed by the band structure calculations, we calculated the optical properties by the all-electron full-potential linearized augmented planewave (LAPW) method[37]. The absorption spectrum of the predicted silicon structures is shown in Fig. 4, the spectra of the diamond silicon and $CuInSe_2$ are also plotted for comparison. The four predicted structures have considerably large absorption than diamond silicon in the visible light range of around 1.7 to 3.2 eV, as plotted in the Fig. 4. This is in agreement with the electronic band structure properties discussed above. Besides, the calculated absorption values of these silicon structures are also comparable to those of $CuInSe_2$ compounds, which are excellent materials for solar cell. Apart from the better absorption in the visible light range, structure *C2/m*-16 has a pretty pronounced absorption peak around 0.5 eV, indicating that this structure may also have potential applications in the infrared range. These calculated absorption spectra indicate that the newly predicted silicon clathrate structures may have promising potential applications in high efficiency solar cells.

In conclusion, to explore silicon clathrates with proper band gaps for photovoltaic applications, we performed crystal structure searches using a two-step method combining with *ab initio* calculations. Inspired by the successful synthesis of silicon clathrate Si-24, we first searched K-Si guest-host clathrate-like compounds under pressure, then removed the potassium and obtained guest-free porous silicon structures. By setting strict criteria including energetics and bandgaps, we picked out four best silicon clathrate structures from thousands of candidates. Formation energies and phonon spectra calculations confirm the stability of these predicted structures with and without potassium guest atoms. These results indicate the possibility to synthesize the clathrate silicon allotropes under pressure with the proper proportion of guest atoms. Our proposed silicon structures have proper bandgaps within the range of the Shockley-Queisser limit. Compared to the conventional diamond silicon, their absorption spectra exhibit a significantly improved overlap with the solar spectrum, thus providing appealing features for applications in solar cells.

Supplementary Material
The details of the crystal structures for all the interested potassium silicon compounds and their phonon spectra are shown in the supplemental material

We thank the financial support provided by the National Key R&D program of China (Grant No: 2016YFA0300404), the National Key Projects for Basic Research in China (Grant No.2015CB921202), the National Natural Science Foundation of China (Grant Nos: 11574133 and 51372112), the NSF Jiangsu province (No. BK20150012), the Science Challenge Project (No. TZ2016001), the Fundamental Research Funds for the Central Universities, and Special Program for Applied Research on Super Computation of the NSFC-Guangdong Joint Fund (the 2nd phase). Calculations were performed on the supercomputer in the HPCC of Nanjing University and "Tianhe-2" at NSCC-Guangzhou.

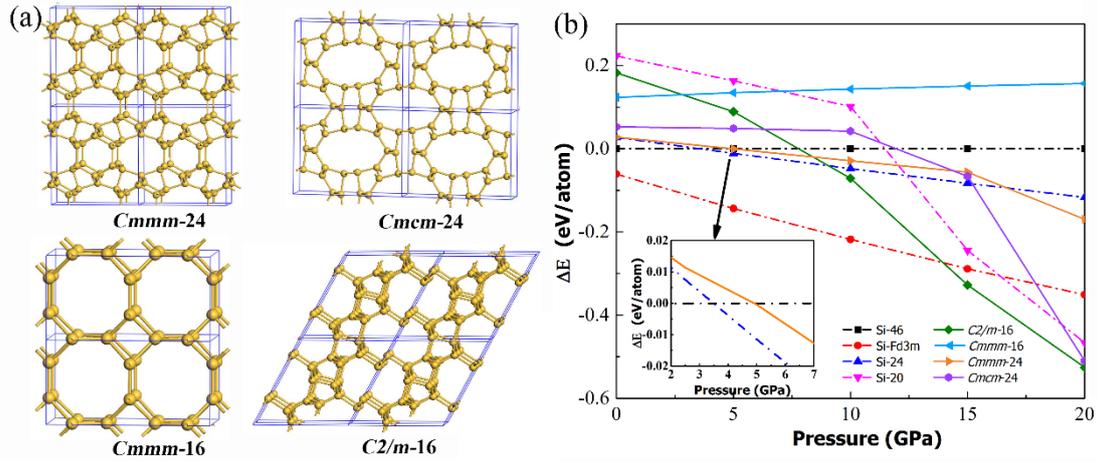

Fig. 1. (a) The crystal structures of the predicted silicon clathrate structures. The structures are labeled by the space group in the Hermann-Mauguin notation together with the number indicating the silicon atoms in the conventional cell. (b) Enthalpy difference relative to Si-46 as a function of pressure. The dashed lines are known structures, and the solid lines are structures proposed in this work. All proposed silicon structures are metastable, but they are more energetically favorable than the theoretically proposed Si-20 structure at ambient pressure. Structure *Cmmm*-24 has a small difference with the synthesized Si-24 and becomes more favorable than Si-46 when the pressure is above 5 GPa, as shown in the inset.

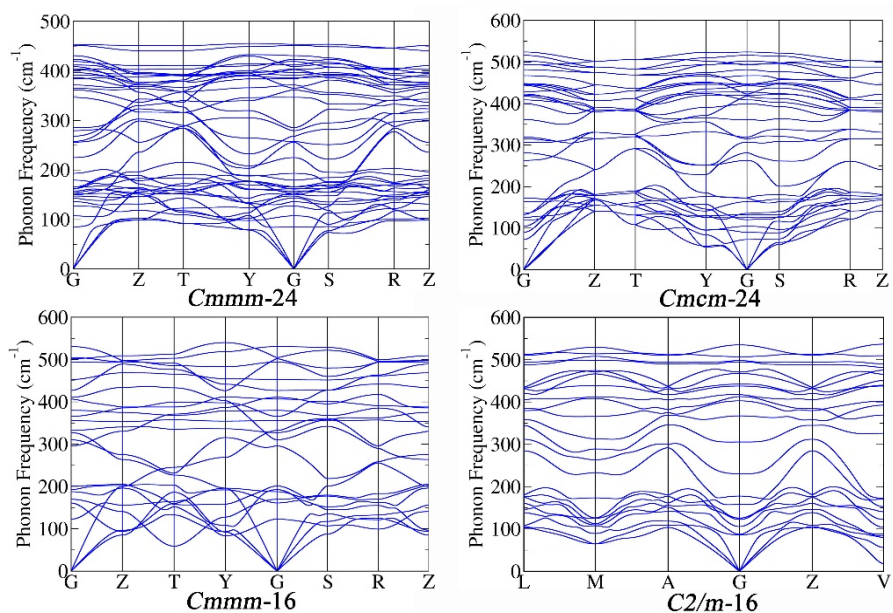

Fig. 2. Phonon spectra of the proposed guest-free clathrate silicon structures at ambient condition. All of them have no imaginary frequencies and suggest that these structures are dynamically stable.

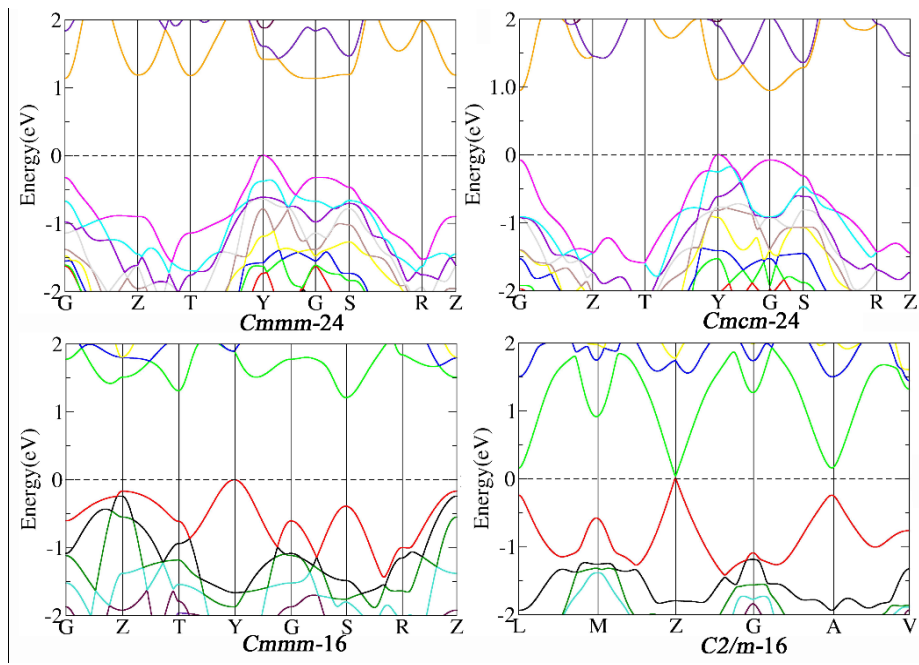

Fig. 3. Electronic band structures of the proposed silicon clathrate structures at ambient pressure.

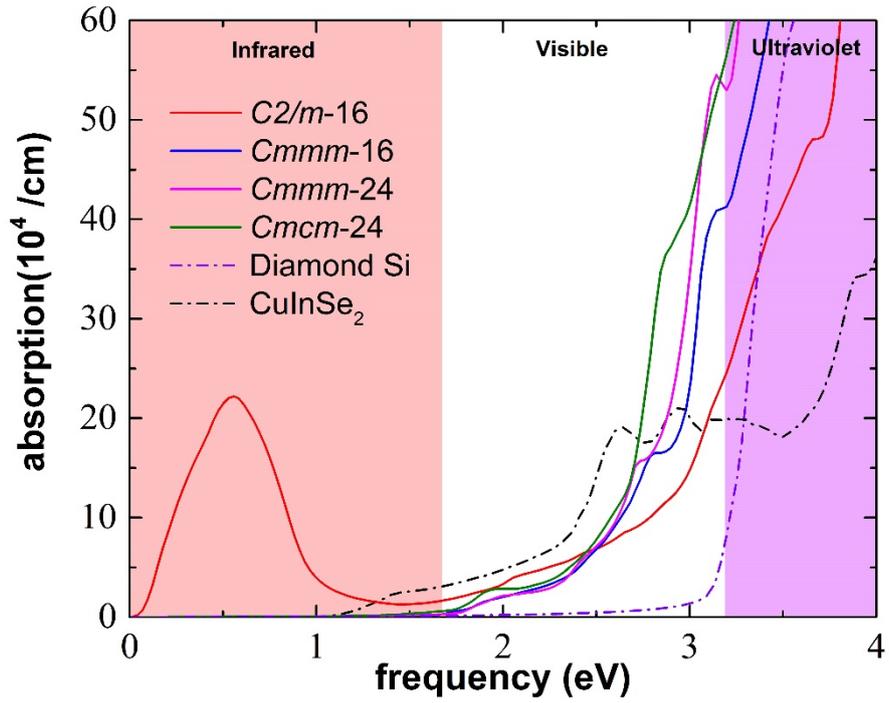

Fig. 4. The absorption spectrum of the proposed silicon structures, diamond silicon and CuInSe$_2$. The curves of diamond silicon and CuInSe$_2$ are plotted with dashed lines. The area between 1.67 eV and 3.19 eV is the energy range of visible sunlight. All the candidates exhibit better absorption spectrum compared to conventional diamond silicon.

Table 1. The formation energy $H_f$ of the compound structures at their stable pressure. The $H_f$ values of the four predicted silicon structures are all negative, indicating the configurations we predicted are energetically stable under high pressure.

| Crystal | C2/m-16 | Cmmm-16 | Cmmm-24 | Cmcm-24 |
|---|---|---|---|---|
| Pressure(GPa) | 20 | 10 | 10 | 20 |
| $H_f$(eV) | -5.50 | -0.027 | -6.71 | -3.94 |

Table 2. Bandgaps of the diamond silicon and proposed silicon clathrate structures. (D) indicates the direct bandgap and the letter is the corresponding position in the Brilliouin zone. (ID) indicates indirect band gap.

| Crystal | C2/m-16 | Cmmm-16 | Cmmm-24 | Cmcm-24 | D-Si |
|---|---|---|---|---|---|
| Band gap (eV) | Z:0.035(D) | Y→S: 1.21(ID) | Y→G: 1.13(ID) | Y→G: 0.95(ID) | G→X: 1.20(ID) |
|  |  | S: 1.60(D) | Y: 1.42(D) | Y: 1.10(D) | G: 3.1(D) |
|  |  | Z: 1.68(D) | G: 1.44(D) | G: 1.02(D) |  |
|  |  | Y: 1.88(D) |  |  |  |